\documentclass[11pt]{article}

\usepackage{a4wide}
\usepackage{latexsym}
\usepackage{graphicx}
\usepackage{color}
\usepackage{hyperref}

\title{A Stochastic Evolutionary Model Exhibiting Power-Law Behaviour
with an Exponential Cutoff}

\author{Trevor Fenner, Mark Levene, and George Loizou \\
School of Computer Science and Information Systems \\
Birkbeck College, University of London \\
London WC1E 7HX, U.K. \\ \{trevor,mark,george\}@dcs.bbk.ac.uk}

\date{}

\begin{document}

\maketitle

\newtheorem{theorem}{Theorem}[section]
\newtheorem{corollary}[theorem]{Corollary}
\newtheorem{lemma}[theorem]{Lemma}
\newtheorem{proposition}[theorem]{Proposition}
\newtheorem{definition}{Definition}[section]
\newtheorem{algorithm}{Algorithm}
\newtheorem{example}{Example}[section]

\begin{abstract}

Recently several authors have proposed stochastic evolutionary models for the growth
of complex networks that give rise to power-law distributions. These models are based
on the notion of preferential attachment leading to the ``rich get richer''
phenomenon. Despite the generality of the proposed stochastic models, there are still
some unexplained phenomena, which may arise due to the limited size of networks such
as protein and e-mail networks. Such networks may in fact exhibit an exponential
cutoff in the power-law scaling, although this cutoff may only be observable in the
tail of the distribution for extremely large networks. We propose a modification of
the basic stochastic evolutionary model, so that after a node is chosen
preferentially, say according to the number of its inlinks, there is a small
probability that this node will be discarded. We show that as a result of this
modification, by viewing the stochastic process in terms of an urn transfer model, we
obtain a power-law distribution with an exponential cutoff. Unlike many other models,
the current model can capture instances where the exponent of the distribution is
less than or equal to two. As a proof of concept, we demonstrate the consistency of
our model by analysing a yeast protein interaction network, the distribution of which
is known to follow a power law with an exponential cutoff.

\end{abstract}

\section{Introduction}

Power-law distributions taking the form
\begin{equation}\label{eq:power-law}
f(i) = C \ i^{- \tau},
\end{equation}
where $C$ and $\tau$ are positive constants, are abundant in nature \cite{SORN00}.
The constant $\tau$ is called the {\em exponent} of the distribution. Examples of
such distributions are: {\em Zipf's law}, which states that the relative frequency of
words in a text is inversely proportional to their rank, {\em Pareto's law}, which
states that the number of people whose personal income is above a certain level
follows a power-law distribution with an exponent between 1.5 and 2 (Pareto's law is
also known as the {\em 80:20 law}, stating that about 20\% of the population earn
80\% of the income) and {\em Gutenberg-Richter's law}, which states that, over a
period of time, the number of earthquakes of a certain magnitude is roughly inversely
proportional to the magnitude. Recently, several researchers have detected power-law
distributions in the topology of various networks such as the World-Wide-Web
\cite{BROD00,KUMA00b} and author citation graphs \cite{REDN98}.

\medskip

The motivation for the current research is two-fold. First, from a complex network
perspective, we would like to understand the stochastic mechanisms that govern the
growth of a network. This has lead to fruitful interdisciplinary research by a
mixture of Computer Scientists, Mathematicians, Statisticians, Physicists, and Social
Scientists \cite{ALBE01,DORO00c,KRAP00,NEWM01,PENN02}, who are actively involved in
investigating various characteristics of complex networks, such as the degree
distribution of the nodes, the diameter, and the relative sizes of various
components. These researchers have proposed several stochastic models for the
evolution of complex networks; all of these have the common theme of {\em
preferential attachment} --- which results in the ``rich get richer'' phenomenon ---
for example, where new links to existing nodes are added in proportion to the number
of links to these nodes currently present. Considering the web as an example of a
complex network, one of the challenges in this line of research is to explain the
empirically discovered power-law distributions \cite{ADAM01b}. It turns out that the
evolutionary model of preferential attachment fails to explain several of the
empirical results, due to the fact that the exponents predicted are inconsistent with
the observations. To address this problem, we proposed in \cite{LEVE01c} an extension
of the stochastic model for the web's evolution in which the addition of links
utilises a mixture of preferential and non-preferential mechanisms. We introduced a
general stochastic model involving the transfer of balls between urns that also
naturally models quantities such as the numbers of web pages in and visitors to a web
site, which are not naturally described in graph-theoretic terms.

Another extension of the preferential attachment model, proposed in \cite{DORO00c},
takes into account the ageing of nodes, so that a link is connected to an old node,
not only preferentially, but also depending on the age of the node: the older the
node is, the less likely it is that other nodes will be connected to it. It was shown
in \cite{DORO00c} that if the ageing function is a power law then the degree
distribution has a phase transition from a power-law distribution, when the exponent
of the ageing function is less than one, to an exponential distribution, when the
exponent is greater than one. A different model of node ageing was proposed in
\cite{AMAR00} with two alternative ageing functions. With the first function the time
a node remains `active', i.e. may acquire new links, decays exponentially, and with
the second function a node remains active until it has acquired a maximum number of
links. Both functions were shown by simulation to lead to an exponential cutoff in
the degree distribution, and for strong enough constraints the distribution appeared
to be purely exponential. Another explanation for a cutoff, offered in \cite{MOSS02},
is that when a link is created the author of the link has limited information
processing capabilities and thus only considers linking to a fraction of the existing
nodes, those that appear to be ``interesting''. It was shown by simulation that when
the fraction of ``interesting nodes'' is less than one there is a change from a
power-law distribution to one that exhibits an exponential cutoff, leading eventually
to an exponential distribution when the fraction is much less than one.

\smallskip

A second motivation for this research is that the viability and efficiency of network
algorithmics are affected by the statistical distributions that govern the network's
structure. For example, the discovered power-law distributions in the web have
recently found applications in local search strategies in web graphs \cite{ADAM01a},
compression of web graphs \cite{ADLE01} and an analysis of the robustness of networks
against error and attack \cite{ALBE00c,JEON01}.

\medskip

Despite the generality of the proposed stochastic models for the evolution of complex
networks, there are still some unexplained phenomena; these may arise due to the
limited size of networks such as protein, e-mail, actor and collaboration networks.
Such networks may in fact exhibit an exponential cutoff in the power-law scaling,
although this cutoff may only be observable in the tail of the distribution for
extremely large networks. The exponential cutoff is of the form
\begin{equation}\label{eq:exp-cutoff}
f(i) = C \ i^{- \tau} q^i,
\end{equation}
with $0 < q < 1$. The exponent $\tau$  in (\ref{eq:exp-cutoff}) will be smaller than
the exponent that would be obtained if we tried to fit a power law without a cutoff,
like (\ref{eq:power-law}), to the data. Unlike many other models leading to power-law
distributions, models with a cutoff can capture situations in which the exponent of
the distribution is less than or equal to two, which would otherwise have infinite
expectation.

\medskip

An exponential cutoff has been observed in protein networks \cite{JEON01}, in e-mail
networks \cite{EBEL02}, in actor networks \cite{AMAR00}, in collaboration networks
\cite{NEWM01,GROSS02b}, and is apparently also present in the distribution of inlinks
in the web graph \cite{MOSS02}, where a cutoff had not previously been observed. We
believe it is likely, in many such cases where power-law distributions have been
observed, that better models would be obtained with an exponential cutoff like
(\ref{eq:exp-cutoff}), with $q$ very close to one.

\medskip

The main aim of this paper is to provide a stochastic evolutionary model that results
in asymptotic power-law distributions with an exponential cutoff, thus allowing us to
model discrete finite systems more accurately and, in addition, enabling us to
explain phenomena where the exponent is less than or equal to two. As with many of
these stochastic growth models, the ideas originated from Simon's visionary paper
published in 1955 \cite{SIMO55}. At the very beginning of his paper, in equation
(1.1), Simon observed that the class of distribution functions he was about to
analyse can be approximated by a distribution like (\ref{eq:exp-cutoff}); he called
the term $q^i$ the {\em convergence factor} and suggested that $q$ is close to one.
He then went on to present his well-known model that yields power-law distributions
like (\ref{eq:power-law}), and which has provided the basis for the models
rediscovered over 40 years later. Simon gave no explanation for the appearance, in
practice, of the convergence factor.

Considering, for example, the web graph, the modification we make to the basic model
to explain the convergence factor is that after a web page is chosen preferentially,
say according to the number of its inlinks, there is a small probability that this
page will be discarded. A possible reason for this may be that the web page has
acquired too many inlinks and therefore needs to be redesigned, or simply that an
error has occurred and the page is lost. Other examples are e-mail networks, where
new users join and old users leave the network, and protein networks, where proteins
may appear or disappear from the network over time.

\smallskip

Networks with an exponential cutoff fall into two categories. The first category of
network, which includes actor and collaboration networks, is monotonically
increasing, i.e. nodes and links are never removed from such networks. In this
category nodes can be either active, in which case they can be the source or
destinations of new links, or inactive in which case they are not involved in any new
links from the time they first become inactive. The second category of network, which
includes the web graph, e-mail and protein networks, is non-monotonic, i.e. links and
nodes may be removed. In this paper we consider the second category of network, but
only allow node deletion. (In \cite{FENN03b} we considered the case where only link
removal is allowed and showed that in that case the degree distribution follows a
power law.) The first category of network (which also exhibits an exponential cutoff
in the degree distribution) will be dealt with in a follow-up paper.

\medskip

The rest of the paper is organised as follows. In Section~\ref{sec:urn} we present an
urn transfer model that extends Simon's model by allowing a ball to sometimes be
discarded. In Section~\ref{sec:power-law} we derive the steady state distribution of
the model, which, as stated, follows an asymptotic power law with an exponential
cutoff like (\ref{eq:exp-cutoff}). In Section~\ref{sec:protein-net} we demonstrate
that our model can provide an explanation of the empirical distributions found in
protein networks. Finally, in Section~\ref{sec:concluding} we give our concluding
remarks.

\section{An Urn Transfer Model}
\label{sec:urn}

We now present an {\em urn transfer model} \cite{JOHN77} for a stochastic process
that emulates the situation when balls (which might represent, for example, proteins
or email accounts) are discarded with a small probability. This model can be viewed
as an extension of Simon's model \cite{SIMO55}, where either a ball is added to the
first urn with probability $p$, or some ball is moved along from the urn it is in to
the next urn with probability $1-p$. We assume that a ball in the $i$th urn has $i$
pins attached to it (which might represent, for example, interactions between
proteins or e-mail messages between email accounts). We note that there is a
correspondence between the Barab\'asi and Albert model \cite{BARA99b}, defined in
terms of nodes and links, and Simon's model, defined in terms of balls and pins, as
was established in \cite{BORN01}. Essentially, the correspondence is obtained by
noting that the balls in an urn can be viewed as an equivalence class of nodes all
having the same connectivity (i.e. degree).

\medskip

We assume a countable number of urns, $urn_1, urn_2, urn_3, \ldots \ $. Initially all
the urns are empty except $urn_1$, which has one ball in it. Let $F_i(k)$ be the
number of balls in $urn_i$ at stage $k$ of the stochastic process, so $F_1(1) = 1$
and all other $F_i(1) = 0$. Then, at stage $k+1$ of the stochastic process, where $k
\ge 1$, one of two things may occur:

\renewcommand{\labelenumi}{(\roman{enumi})}
\begin{enumerate}
\item with probability $p$, $0 < p < 1$, a new ball (with one pin attached) is
inserted into $urn_1$, or

\item with probability $1 - p$ an urn is selected, with $urn_i$ being selected with
probability proportional to $i F_i(k)$ (i.e. $urn_i$ is selected preferentially in
proportion to the total number of pins it contains), and a ball is chosen from the
selected urn, $urn_i$ say; then,

\begin{enumerate}
\item with probability $q$, $0 < q \le 1$, the chosen ball is transferred to
$urn_{i+1}$, (this is equivalent to attaching an additional pin to the ball chosen
from $urn_i$), or

\item with probability $1 - q$ the ball chosen is discarded.
\end{enumerate}
\end{enumerate}
\smallskip

The expected total number of balls in the urns  at stage $k$ is given by
\begin{eqnarray}\label{eq:pins2}
E \Big( \sum_{i=1}^k F_i(k) \Big) & = & 1 + (k - 1) \Big(p - (1 - p)
(1 - q) \Big) \nonumber \\
& = & (1-p) (2-q) + k  \Big(p - (1 - p) (1 - q) \Big).
\end{eqnarray}
\medskip

We note that we could modify the initial conditions so that, for example, $urn_1$
initially contained $\delta > 1$ balls instead of one. It can be shown, from the
development of the model below, that any change in the initial conditions will have
no effect on the asymptotic distribution of the balls in the urns as $k$ tends to
infinity, provided the process does not terminate with all of the urns empty.

\medskip

To ensure that, on average, more balls are added to the system than are discarded, on
account of (\ref{eq:pins2}) we require $p > (1-p) (1-q)$, which implies
\begin{displaymath}\label{eq:q}
q > \frac{1 - 2 p}{1 - p};
\end{displaymath}
this trivially holds for $p > 1/2$.

\smallskip

From now on we assume that this holds. This constraint implies that the probability
that the urn transfer process will {\em not} terminate with all the urns being empty
is positive. More specifically, the probability of non-termination is given by
\begin{equation}\label{eq:gambler}
1 - \left( \frac{(1-p) (1-q)}{p} \right)^\delta,
\end{equation}
where $\delta$ is the initial number of balls in $urn_1$; this is exactly the
probability that the gambler's fortune will increase forever \cite{ROSS83}.

\medskip

The total number of pins attached to balls in $urn_i$ at stage $k$ is $i F_i(k)$, so
the expected total number of pins in the urns is given by
\begin{eqnarray}\label{eq:pins1}
E \Big( \sum_{i=1}^k i F_i(k) \Big) & = & 1 + (k - 1) \Big(p +
(1-p) q \Big) - (1-p) (1-q) \sum_{j=1}^{k-1} \theta_j  \nonumber \\
& = & k \Big( p + (1-p) q \Big) - (1-p) (1-q) \Big( \sum_{j=1}^{k-1} \theta_j -1
\Big),
\end{eqnarray}
where $\theta_j$, $1 \le j \le k-1$, is the expectation of $\Theta_j$, the number of
pins attached to the ball chosen at step (ii) of stage $j$ (i.e. the urn number). So
\begin{equation}\label{eq:theta-define}
\theta_j = E(\Theta_j) = E \left( \frac{\sum_{i=1}^j i^2 F_i(j)}{\sum_{i=1}^j i
F_i(j)} \right).
\end{equation}
\smallskip

\noindent As a consequence we have
\begin{displaymath}\label{eq:thetaj-bounds}
1 \le \theta_j \le j,
\end{displaymath}
since at stage $j$ there cannot be more than $j$ pins in the system.


\medskip

Now let
\begin{displaymath}\label{eq:thetaj}
\theta^{(k)} = \frac{1}{k} \sum_{j=1}^k \theta_j.
\end{displaymath}
\smallskip

\noindent Since there are at least as many pins in the system as there are balls, it
follows from (\ref{eq:pins2}) and (\ref{eq:pins1}) that
\begin{equation}\label{eq:thetak-bounds}
1 \le \theta^{(k)} \le \frac{1}{1-q}.
\end{equation}
\smallskip

So, since $\theta^{(k)}$ is bounded, we will make the reasonable assumption that
$\theta^{(k)}$ tends to a limit $\theta$ as $k$ tends to infinity, i.e.
\begin{displaymath}\label{eq:theta-limit}
\lim_{k \to \infty} \theta^{(k)}= \theta.
\end{displaymath}
\smallskip

Letting $k$ tend to infinity in (\ref{eq:thetak-bounds}) gives
\begin{displaymath}\label{eq:theta-bounds}
1 \le \theta \le \frac{1}{1-q}.
\end{displaymath}
\medskip

In the next section we demonstrate through simulation of the stochastic process that
our assumption that $\theta^{(k)}$ converges appears to hold. We also explain how the
asymptotic value $\theta$ may be obtained, assuming that the limit exists.

\section{Derivation of the Steady State Distribution}
\label{sec:power-law}

Following Simon \cite{SIMO55}, we now state the mean-field equations for the urn
transfer model. For $i > 1$ we have
\begin{equation}\label{eq:ss0}
E_k(F_i(k+1)) = F_i(k) + \beta_k \Big(q (i-1) F_{i-1}(k) - i F_i(k) \Big),
\end{equation}
where $E_k(F_i(k+1))$ is the expected value of $F_i(k+1)$ given the state of the
model at stage $k$, and
\begin{equation}\label{eq:betak}
\beta_k = \frac{1 - p}{\sum_{i=1}^k \ i F_i(k)}
\end{equation}
is the normalising factor.

\medskip

Equation (\ref{eq:ss0}) gives the expected number of balls in $urn_i$ at stage $k+1$.
This is equal to the previous number of balls in $urn_i$ plus the probability of
adding a ball to $urn_i$ minus the probability of removing a ball from $urn_i$. The
former probability is just the probability of choosing a ball from $urn_{i-1}$ and
transferring it to $urn_i$ in step (ii)(a) of the stochastic process defined in
Section~\ref{sec:urn}, whilst the latter probability is the probability of choosing a
ball from $urn_i$ in step (ii) of the process.

\medskip

In the boundary case, $i = 1$, we have
\begin{equation}\label{eq:initial}
E_k(F_1(k+1)) = F_1(k) + p - \beta_k \ F_1(k),
\end{equation}
\smallskip
for the expected number of balls in $urn_1$, which is equal to the previous number of
balls in the first urn plus the probability of inserting a new ball into $urn_1$ in
step (i) of the stochastic process defined in Section~\ref{sec:urn} minus the
probability of choosing a ball from $urn_1$ in step (ii).

\medskip

In order to solve the equations for the model, we make the assumption that, for large
$k$, the random variable $\beta_k$ can be approximated by a constant (i.e.
non-random) value depending only on $k$. We take this approximation to be
\begin{displaymath}
\hat{\beta}_k = \frac{1-p}{k \left( p + (1-p) q - (1-p) (1-q) \theta^{(k)} \right)}.
\end{displaymath}
\smallskip

The motivation for this approximation is that the denominator in the definition of
$\beta_k$ has been replaced by an asymptotic approximation of its expectation as
given in (\ref{eq:pins1}). We observe that replacing $\beta_k$ by $\hat{\beta}_k$
results in an approximation similar to that of the ``$p_k$ model'' in \cite{LEVE01c},
which is essentially a {\em mean-field} approach.

\medskip

We can now take expectations of (\ref{eq:ss0}) and (\ref{eq:initial}). Thus, by the
linearity of the expectation operator $E(.)$, we obtain
\begin{equation}\label{eq:expected-ss0}
E(F_i(k+1)) = E(F_i(k)) + \hat{\beta}_k \Big( q (i-1) E(F_{i-1}(k)) - i E(F_i(k))
\Big)
\end{equation}
and
\begin{equation}\label{eq:expected-initial}
E(F_1(k+1)) = E(F_1(k)) + p - \hat{\beta}_k E(F_1(k)).
\end{equation}
\medskip

In order to obtain an asymptotic solution of equations (\ref{eq:expected-ss0}) and
(\ref{eq:expected-initial}), we require that $E(F_i (k)) / k$ converges to $f_i$ as
$k$ tends to infinity. Suppose for the moment that this is the case, then, provided
the convergence is fast enough, $E(F_i(k+1)) - E(F_i(k))$ tends to $f_i$. By ``fast
enough'' we mean $\epsilon_{i,k+1} - \epsilon_{i,k}$ is $o(1/k)$ for large $k$, where
\begin{displaymath}
E(F_i(k)) = k (f_i + \epsilon_{i,k}).
\end{displaymath}
\smallskip

Now, letting
\begin{equation}\label{eq:beta}
\beta = \frac{1-p}{p + (1-p) q - (1-p) (1-q) \theta},
\end{equation}
we see that $\hat{\beta_k} E(F_i(k))$ tends to $\beta f_i$ as $k$ tends to infinity.

\medskip

So, letting $k$ tend to infinity, (\ref{eq:expected-ss0}) and
(\ref{eq:expected-initial}) yield
\begin{equation}\label{eq:ss2}
f_i = \beta \Big(q (i-1) f_{i-1} - i f_i \Big),
\end{equation}
for $i > 1$, and
\begin{equation}\label{eq:ss1}
f_1 = p - \beta f_1.
\end{equation}
\medskip


Following the lines of the proof given in the Appendix of \cite{LEVE01c}, we can show
that $\epsilon_{i,k}$ tends to zero as $k$ tends to infinity provided we make the
further assumption that
\begin{displaymath}
\mid \! \theta - \theta^{(k)} \! \mid \le \frac{c}{k},
\end{displaymath}
for some constant $c$. In other words, this assumption states that the expected
number of pins attached to the balls chosen in the first $k$ stages of the stochastic
process is within a constant of the asymptotic expected number of pins attached to
the chosen ball multiplied by $k$, i.e.
\begin{displaymath}
k \theta - c \le \sum_{j=1}^k \theta_j \le k \theta + c.
\end{displaymath}
\smallskip

In order to verify the convergence we ran some simulations; these will be discussed
at the end of this section.

\medskip

Provided that $\beta_k$ can be approximated by $\hat{\beta}_k$ for large $k$, then,
under the stated assumptions, $f_i$ is the asymptotic expected rate of increase of
the number of balls in $urn_i$; thus the asymptotic proportion of balls in $urn_i$ is
proportional to $f_i$.

\medskip

From (\ref{eq:ss2}) and (\ref{eq:ss1}) we obtain
\begin{equation}\label{eq:ss4}
f_i = \frac{\beta \ q \ (i-1)}{1 + i \beta} \ f_{i-1} = \frac{q \ (i-1)}{i  +
\varrho} \ f_{i-1}
\end{equation}
and
\begin{equation}\label{eq:ss3}
f_1 = \frac{p}{1 + \beta} = \frac{\varrho \ p}{1 + \varrho},
\end{equation}
where $\varrho = 1 / \beta$. Now, on repeatedly using (\ref{eq:ss4}), we get
\begin{equation}\label{eq:ss5}
f_i = \frac{\varrho \ p \ q^{i-1} \ 1 \ 2 \ \cdots \ (i-1)}{(1 + \varrho) \ (2 +
\varrho) \ \cdots \ (i + \varrho)} = \frac{\varrho \ p \ \Gamma(1 + \varrho) \
\Gamma(i) \ q^i}{q \ \Gamma(i + 1 + \varrho)},
\end{equation}
where $\Gamma$ is the gamma function \cite[6.1]{ABRA72}.

\medskip

Thus for large $i$, on using Stirling's approximation \cite[6.1.39]{ABRA72}, we
obtain $f_i$ in a form corresponding to (\ref{eq:exp-cutoff}):
\begin{equation}\label{eq:plc}
f_i \sim \frac{C \ q^i}{i^{1 + \varrho}},
\end{equation}
where $\sim$ means {\em is asymptotic to}, and
\begin{displaymath}
C = \frac{\varrho \ p \ \Gamma(1 + \varrho)}{q}.
\end{displaymath}
\smallskip

From (\ref{eq:ss5}), it follows that
\begin{eqnarray}\label{eq:hyperg1}
\sum_{i=1}^\infty f_i & = & \frac{\varrho \ p \ \Gamma(2 + \varrho)}{1 + \varrho} \
\sum_{j=0}^\infty
\frac{\Gamma(j + 1) \ \Gamma(j + 1)}{\Gamma(j + 2 + \varrho)} \frac{q^{j}}{j !} \nonumber \\
& = & \frac{\varrho \ p}{1 + \varrho} \ F(1, 1; 2 + \varrho; q),
\end{eqnarray}
where $F$ is the hypergeometric function \cite[15.1.1]{ABRA72}. From
(\ref{eq:hyperg1}) it is immediate that the first moment is given by
\begin{equation}\label{eq:hyperg2}
\sum_{i=1}^\infty i f_i = \frac{\varrho \ p}{1 + \varrho} \ F(1, 2; 2 + \varrho; q),
\end{equation}
and the second moment is given by
\begin{equation}\label{eq:hyperg3}
\sum_{i=1}^\infty i^2 f_i = \frac{\varrho \ p}{1 + \varrho} \ F(2, 2; 2 + \varrho;
q).
\end{equation}
\medskip


Under the assumptions we have made for the steady state distribution, using
(\ref{eq:theta-define}), (\ref{eq:hyperg2}) and (\ref{eq:hyperg3}) we obtain
\begin{equation}\label{eq:hyperg-theta}
\theta = \frac{F(2, 2; 2 + \varrho; q)}{F(1, 2; 2 + \varrho; q)}.
\end{equation}
\medskip

In the special case when $q = 1$, which is Simon's original model, using the fact
that in this case $\varrho = 1/(1-p)$, we obtain by Gauss's formula
\cite[15.1.20]{ABRA72}
\begin{displaymath}
\sum_{i=1}^\infty i f_i = 1,
\end{displaymath}
as expected, and
\begin{displaymath}
\sum_{i=1}^\infty i^2 f_i = \frac{1}{2 p - 1},
\end{displaymath}
which is valid only if $p > 0.5$, i.e. $\varrho > 2$.

\medskip

Letting $k$ tend to infinity in (\ref{eq:pins1}) and using (\ref{eq:beta}) we obtain
\begin{displaymath}\label{eq:sum-fi}
\sum_{i=1}^\infty i f_i = \frac{1-p}{\beta} = \varrho (1-p).
\end{displaymath}
\smallskip

Together with (\ref{eq:hyperg2}), this gives the following equation for $\varrho$ in
terms of $p$ and $q$:
\begin{equation}\label{eq:solve-pins2}
(1-p) (1 + \varrho) =  p \ F(1, 2; 2 + \varrho; q).
\end{equation}
\smallskip

This equation may be solved numerically to obtain the value of $\varrho$, and
$\theta$ can then be obtained from (\ref{eq:beta}) or (\ref{eq:hyperg-theta}). (It
can be shown that, by virtue of (\ref{eq:solve-pins2}), both equations yield the same
value for $\theta$.)

\medskip

In order to verify the convergence assumptions we ran several stochastic simulations.
For $p = 0.3$ and $q = 0.975$, the number of aborts (i.e. computations terminating
because all the urns were empty) predicted by (\ref{eq:gambler}) is about 5.8\%. We
ran the simulations 1000 times, each run for 1000 steps. There were 49 aborts, with
the maximum number of steps before aborting being 12 and the average 4.14.

\smallskip

\begin{figure}[ht]
\centerline{\includegraphics[width=12cm,height=9.33cm]{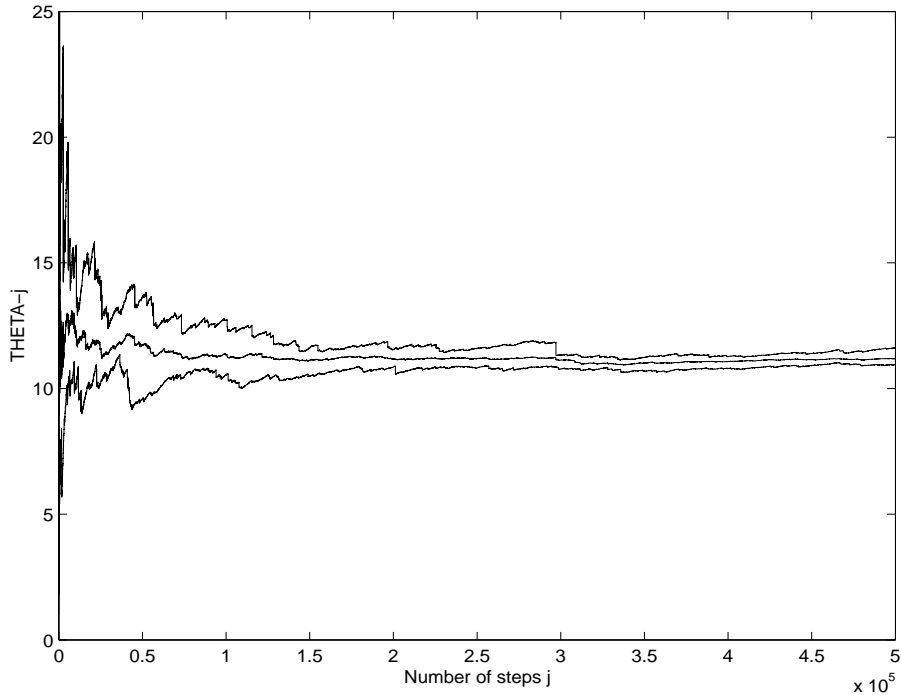}}
\caption{\label{fig:converge} Convergence of $\Theta_j$ to $\theta$}
\end{figure}

Figure~\ref{fig:converge} shows a summary of ten runs, each of half a million steps,
with the above parameters, plotting $\Theta_j$ against $j$. The bottom plot is the
minimum $\Theta_j$ over the ten runs, the middle plot is the average and the top plot
is the maximum. The asymptotic average value of $\Theta_j$ was $11.1902$. On the
other hand, computing $\theta$ from (\ref{eq:hyperg-theta}), averaged over ten runs,
with $\varrho$ taken to be $(k \beta_k)^{-1}$, gave 11.1762. As a final check,
computing $\varrho$ from (\ref{eq:solve-pins2}) and $\theta$ from
(\ref{eq:hyperg-theta}) or (\ref{eq:beta}) gives 11.1753. In further simulations,
similar plots were obtained for other values of $p$ and $q$.

\medskip


\begin{figure}[ht]
\centerline{\includegraphics[width=12cm,height=9.33cm]{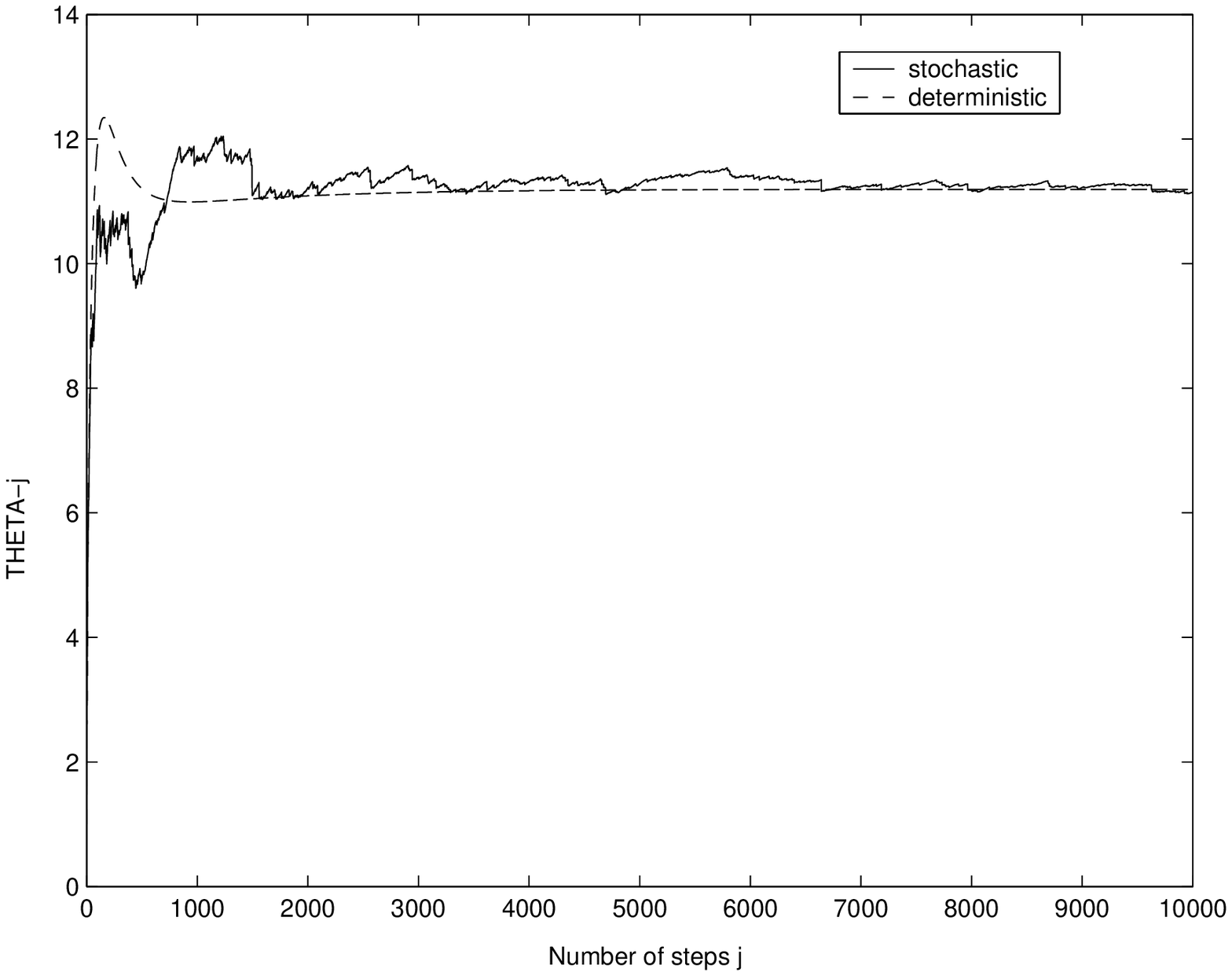}}
\caption{\label{fig:compare} Comparison of the stochastic simulation and the
deterministic computation}
\end{figure}
\medskip

As a further validation, we ran the stochastic simulation for 10,000 steps, repeated
100 times, with parameters $p = 0.3$ and $q = 0.975$ as above, and compared it with a
deterministic computation of the process using (\ref{eq:ss0}), (\ref{eq:betak}) and
(\ref{eq:initial}) to calculate $E(F_i(k))$, the expected number of balls in each
urn, instead of $F_i(k)$, the actual number of balls. The expected values will not in
general be integral. The plots of the average $\Theta_j$ against $j$ for the
stochastic simulation (solid line) and the deterministic computation (dashed line)
are shown in Figure~\ref{fig:compare}. The asymptotic average value of $\Theta_j$ for
the deterministic computation was $11.192$ and for the stochastic simulation
$11.145$.

\section{A Model for Protein Networks}
\label{sec:protein-net}

As mentioned in the introduction, exponential cutoff has been observed in several
networks. Our model can be directly applied to web graphs \cite{MOSS02}, where balls
represent web pages and pins represent links, to e-mail networks \cite{EBEL02}, where
balls represent e-mail accounts and pins represent e-mail messages, and to protein
networks \cite{JEON01}, where balls represent proteins and pins represent protein
interactions. In a web graph removing a ball corresponds to deleting a web page, in
an e-mail network removing a ball corresponds to a user's e-mail account being
removed from the network, and in a protein network removing a ball corresponds to
gene loss resulting in the loss of a protein. The other category of network
exhibiting exponential cutoff mentioned in the introduction, such as collaboration
and actor networks, will be the subject of a follow-up paper.

\smallskip

We note that some types of network, viz. protein, collaboration and actor networks,
are essentially undirected. Consequently, in our model, a new interaction between two
proteins, for example, ought to be represented by two separate events, corresponding
to the new interactions for each of the two proteins. This would correspond to taking
in pairs the events of attaching a pin to a ball. We ignore this complication, but
note that many of the models proposed, for example for the web graph, similarly
ignore the difference between directed and undirected graphs (e.g. \cite{BARA99b}).

\smallskip

As a proof of concept, we focus here on protein networks, and in particular we will
examine a yeast protein interaction network \cite{JEON01}. This is an undirected
graph that can be downloaded from \url{www.nd.edu/~networks/database/protein}. To
obtain the values for $\varrho$ and $q$ we performed a nonlinear regression on a
log-log transformation of the degree distribution of the yeast network data obtained
from this website, fitting to the equation
\begin{equation}\label{eq:regress}
y = a - (\varrho+1) x + \ln(q) \exp(x)
\end{equation}
implied by (\ref{eq:plc}), where $a$ is a constant. The values of $\varrho$ and $q$
obtained from the regression of the complete yeast data set are $\varrho = 1.065$ and
$q = 0.9642$.

\smallskip

We next performed a stochastic simulation to test the validity of our model. In order
to carry out the simulation we require values of $p$ and $k$. From (\ref{eq:pins2})
and (\ref{eq:betak}) using the fact that $\varrho \approx (k \beta_k)^{-1}$, we
obtain
\begin{equation}\label{eq:solve-balls}
\frac{balls_k}{k} \approx p - (1-p) (1-q),
\end{equation}
and
\begin{equation}\label{eq:solve-pins}
\frac{pins_k}{k} \approx \varrho (1-p),
\end{equation}
where $balls_k$ and $pins_k$ stand for the expected numbers of balls and pins in the
urns at stage $k$, respectively. (The right-hand sides of (\ref{eq:solve-balls}) and
(\ref{eq:solve-pins}) are the limiting values of the left-hand sides as $k$ tends to
infinity.)

\smallskip

From these we obtain an equation for the branching factor $bf$, viz.
\begin{equation}\label{eq:br-factor}
bf = \frac{pins_k}{balls_k} \approx \frac{\varrho (1-p)}{p - (1-p) (1-q)},
\end{equation}
and from (\ref{eq:br-factor}) we can derive
\begin{equation}\label{eq:p-data}
p \approx \frac{\varrho + bf (1-q)}{\varrho + bf (2-q)}.
\end{equation}
\smallskip

From the original yeast data, the values of $balls_k$ and $pins_k$ were $1870$ and
$4480$, respectively, which give a branching factor  $bf = 2.3957$. Using the values
of $\varrho$ and $q$ from the regression and this value of $bf$, from
(\ref{eq:p-data}) we obtain a value of $p = 0.3245$ to use in the simulation. Using
this value of $p$, from (\ref{eq:solve-balls}) or (\ref{eq:solve-pins}) we obtain a
value of $k = 6227$. (Alternatively, we could have used (\ref{eq:solve-pins2}) to
estimate $p$, giving the value $0.3026$. We preferred the previous method, as this
avoids the sensitivity of the hypergeometric function for values of $q$ near $1$.)

\smallskip

We carried out 10 simulation runs of the stochastic process using these values of
$p$, $q$ and $k$. Using the value of $pins_k$ from the simulations we obtained an
estimate of $\varrho$ from (\ref{eq:solve-pins}). The average value for $balls_k$ was
$1869$, for $pins_k$ was $4795$, and for $\varrho$ was $1.14$.

\smallskip

Although these values seem to provide a good fit to the original data, as a further
validation we investigated the value of $\theta$. Its value can estimated from
(\ref{eq:theta-define}) as
\begin{equation}\label{eq:sqpins}
\theta \approx \frac{sqpins_k}{pins_k},
\end{equation}
where $sqpins_k$ is given by
\begin{displaymath}
sqpins_k = \sum_{i=1}^k i^2 F_i(k).
\end{displaymath}
\smallskip

The value $sqpins_k$ from the original data is $29140$. Using (\ref{eq:sqpins}) this
gives the empirical value $\theta = 6.5045$.

\medskip

We have three equations for $\theta$: the first is the approximation given by
(\ref{eq:sqpins}), the second is (\ref{eq:hyperg-theta}), and the third, derived from
(\ref{eq:beta}), is
\begin{equation}\label{eq:theta3}
\theta = \frac{1 - \varrho (1-p)}{(1-p) (1-q)} - 1,
\end{equation}
remembering that $\varrho = 1/\beta$.

\smallskip

Using the value $sqpins_k = 40195$ from the simulation, the first estimate of
$\theta$, from (\ref{eq:sqpins}), is $8.3823$. The second estimate, from
(\ref{eq:hyperg-theta}), is $9.3057$, while the third estimate, from
(\ref{eq:theta3}), is $10.0629$. It can be seen that the empirical value of $\theta$
is not consistent with these estimates from the mean-field equations.

\medskip

We suggest that one reason for this inconsistency is due to problems in fitting
power-law type distributions, due to difficulties with non-monotonic fluctuations in
the tail. (Another reason maybe the sensitivity of the nonlinear regression to the
cutoff parameter $q$.) In particular, the presence of {\em gaps} in the degree
distribution is the main manifestation of this problem. There is a {\em gap} in the
degree distribution at $i$ if there are no nodes of degree $i$ but there exists some
node of degree $j$, where $j > i$. We discussed this problem more fully in the
context of a pure power-law distribution in \cite{FENN03b}, and concluded that a
preferable approach is to ignore all data points from the first gap onwards. Evidence
of the advantage of discarding data points in the tail of the distribution was also
given in \cite{GOLD04}, who suggest the more radical approach of using only the first
five data points.

\medskip

\begin{figure}[ht]
\centerline{\includegraphics[width=12cm,height=9.33cm]{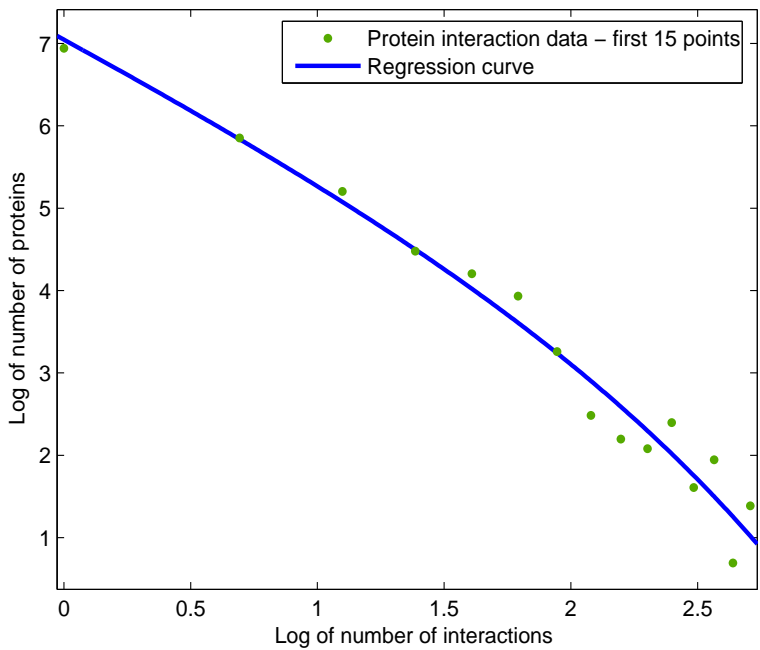}}
\caption{\label{fig:regress} Yeast protein interaction network data}
\end{figure}

Following this approach we only use the first 15 data points in the degree
distribution, since the first gap occurs at $i = 16$. The values of $\varrho$ and $q$
obtained from the nonlinear regression fitting (\ref{eq:regress}) to the truncated
data set, are $\varrho = 0.5586$ and $q = 0.879$. The first 15 data points as well as
the computed regression curve are shown in Figure~\ref{fig:regress}. Using these
values of $\varrho$ and $q$ together with the values of $balls_k$ and $pins_k$ from
the original data, we computed $p = 0.2615$ from (\ref{eq:p-data}) and $k = 10860$
from (\ref{eq:solve-balls}) or (\ref{eq:solve-pins}).

\smallskip

We then repeated the 10 stochastic simulations using the new values of $p, q$ and
$k$. The average estimate for $balls_k$ was $1858$, for $pins_k$ was $4463$, and for
$\varrho$ using the values of $pins_k$ from the simulations and (\ref{eq:solve-pins})
was $0.5565$. The first estimate of $\theta$, using (\ref{eq:sqpins}) and the value
$sqpins_k = 24666$ from the simulation, was $5.5263$. The second estimate was
$5.5660$ from (\ref{eq:hyperg-theta}), while the third estimate was $5.5743$ from
(\ref{eq:theta3}). It can be seen that these estimates are significantly more
consistent with the empirical values of $sqpins_k$ and $\theta$ from the original
data set.

\medskip

We then carried out a further verification of the applicability of our methodology.
We computed the average number of balls in each urn over the second 10 simulation
runs; the first urn that contained no balls in any of the simulations was urn 32.
Next, we performed a nonlinear regression on the first 31 urn averages using
(\ref{eq:regress}), as before. The values of $\varrho$ and $q$ obtained from this
regression were $\varrho = 0.5624$ and $q = 0.8880$, which are very close to the
corresponding values obtained from the regression on the truncated data set.

\medskip

Jeong et al. \cite{JEON01} fit the data set to a power-law distribution with an
exponential cutoff. However, they shift all the degree values by one, in order to
obtain a better fit for small degrees. They report $q$ as approximately $\exp(-0.05)
= 0.9512$ and $\varrho$ as approximately $1.4$ (see supplementary information
provided by the authors of \cite{JEON01}). In order to compare our results with
theirs, we repeated the nonlinear regression on the first 15 data points, using
(\ref{eq:regress}), taking the degrees of the data points to be from 2 to 16. This
gave the values $q = 0.9828$ and $\varrho = 1.513$, which are comparable to Jeong et
al.'s results.

\medskip

A confirmation of the existence of a cutoff can be obtained as follows. Let us assume
there is no cutoff, i.e. $q = 1$. Then, from (\ref{eq:beta}) we have $\varrho =
1/(1-p)$. Using this and (\ref{eq:br-factor}) we derive $bf \approx \varrho /
(\varrho -1)$, and thus $\varrho \approx bf / (bf-1)$. Now, since we are assuming
there is no cutoff, we can perform a linear regression on the the log-log
transformation of the first 15 data points, i.e. fitting to (\ref{eq:regress}) with
$q=1$, which yields $\varrho = 1.251$.

\smallskip

Now, using the empirical branching factor $bf = 2.3957$ from the original data, we
can compute $\varrho$ as $2.3957/1.3957 = 1.7165$. This significantly deviates from
the value $1.251$, obtained from the linear regression. Alternatively, we can compute
the branching factor as $bf \approx 1.251/0.251 = 4.9841$, which deviates
significantly from the empirical branching factor. These discrepancies lead us to
conclude that a cutoff does exist, i.e. that $q < 1$.

\section{Concluding Remarks}
\label{sec:concluding}

We have presented an extension of Simon's classical stochastic process, which results
in a power-law distribution with an exponential cutoff, and for which the power-law
exponent need not exceed two. When viewing this stochastic process in terms of an urn
transfer model, the difference from the classical process is that, after a ball is
chosen on the basis of preferential attachment, with probability $1-q$ the ball is
discarded. Under our assumption that, for large $k$, the normalising factor $\beta_k$
can be approximated by the constant $\hat{\beta}_k$, we have derived the asymptotic
formula (\ref{eq:plc}), which shows that the distribution of the number of balls in
the urns approximately follows a power-law distribution with an exponential cutoff.
We note that we have, in fact, derived a more accurate formula (\ref{eq:ss5}) in
terms of gamma functions.

\smallskip

Exponential cutoffs have been identified in protein, e-mail, actor and collaboration
networks, and possibly in the web graph \cite{MOSS02}; it is likely exponential
cutoffs also occur in other complex networks. Our model assumes that balls are
discarded rather than just becoming inactive as in actor and collaboration networks
(the treatment of such networks with inactive nodes will be dealt in a subsequent
paper). We have validated our model with data from a yeast protein network, showing
that our model provides a possible explanation for the exponential cutoff. We have
also presented convincing numerical evidence of the existence of a cutoff in this
network.

\smallskip

In addition, we have checked that our model is consistent with the emergence of the
power-law distribution for inlinks in the web graph. However, in this case $q$ is
probably very close to one \cite{MOSS02}, and this may be the reason that we have not
managed to establish the existence of an exponential cutoff for the web graph.

\newcommand{\etalchar}[1]{$^{#1}$}


\begin{thebibliography}{BKM{\etalchar{+}}00}

\bibitem[AB02]{ALBE01}
R.~Albert and A.-L. Barab\'asi.
\newblock Statistical mechanics of complex networks.
\newblock {\em Reviews of Modern Physics}, 74:47--97, 2002.

\bibitem[AH01]{ADAM01b}
L.A. Adamic and B.A. Huberman.
\newblock The {W}eb's hidden order.
\newblock {\em Communications of the ACM}, 44(9):55--59, 2001.

\bibitem[AJB00]{ALBE00c}
R.~Albert, H.~Jeong, and A.-L. Barab\'asi.
\newblock Error and attack tolerance of complex networks.
\newblock {\em Nature}, 406:378--382, 2000.

\bibitem[ALPH01]{ADAM01a}
L.A. Adamic, R.M. Lukose, A.R. Puniyani, and B.A. Huberman.
\newblock Search in power-law networks.
\newblock {\em Physical Review E}, 64:046135--1--046135--8, 2001.

\bibitem[AM01]{ADLE01}
M.~Adler and M.~Mitzenmacher.
\newblock Towards compressing web graphs.
\newblock In {\em Proceedings of IEEE Data Compression Conference}, pages
  203--212, Snowbird, Utah, 2001.

\bibitem[AS72]{ABRA72}
M.~Abramowitz and I.A. Stegun, editors.
\newblock {\em Handbook of Mathematical Functions with Formulas, Graphs and
  Mathematical Tables}.
\newblock Dover, New York, NY, 1972.

\bibitem[ASBS00]{AMAR00}
L.A.N. Amaral, A.~Scala, M.~Barth\'el\'emy, and H.E. Stanley.
\newblock Classes of small-world networks.
\newblock {\em Proceedings of the National Academy of Sciences of the United
  States of America}, 97:11149--11152, 2000.

\bibitem[BA99]{BARA99b}
A.-L. Barab\'asi and R.~Albert.
\newblock Emergence of scaling in random networks.
\newblock {\em Science}, 286:509--512, 1999.

\bibitem[BE01]{BORN01}
S.~Bornholdt and H.~Ebel.
\newblock World {W}ide {W}eb scaling exponent from {S}imon's 1955 model.
\newblock {\em Physical Review E}, 64:035104--1--035104--4, 2001.

\bibitem[BKM{\etalchar{+}}00]{BROD00}
A.~Broder, R.~Kumar, F.~Maghoul, P.~Raghavan, A.~Rajagopalan, R.~Stata,
  A.~Tomkins, and J.~Wiener.
\newblock Graph structure in the {W}eb.
\newblock {\em Computer Networks}, 33:309--320, 2000.

\bibitem[DM00]{DORO00c}
S.N. Dorogovtsev and J.F.F. Mendes.
\newblock Evolution of networks with aging of sites.
\newblock {\em Physical Review E}, 62:1842--1845, 2000.

\bibitem[EMB02]{EBEL02}
H.~Ebel, L.-I. Mielsch, and S.~Bornholdt.
\newblock Scale-free topology of e-mail networks.
\newblock {\em Physical Review E}, 66:035103--1--035103--4, 2002.

\bibitem[FLL05]{FENN03b}
T.I. Fenner, M.~Levene, and G.~Loizou.
\newblock A stochastic model for the evolution of the web allowing link
  deletion.
\newblock {\em ACM Transactions on Internet Technology}, August, 2005.
\newblock To appear; also appears in the Condensed Matter Archive,
  cond-mat/0304316.

\bibitem[GMY04]{GOLD04}
M.L. Goldstein, S.A. Morris, and G.G. Yen.
\newblock Problems with fitting to the power-law distribution.
\newblock {\em Condensed Matter Archive}, cond-mat/0402322, 2004.

\bibitem[Gro02]{GROSS02b}
J.W. Grossman.
\newblock Patterns of collaboration in mathematical research.
\newblock {\em SIAM News}, 35(9), November 2002.

\bibitem[JK77]{JOHN77}
N.L. Johnson and S.~Kotz.
\newblock {\em Urn Models and their Application: An Approach to Modern Discrete
  Probability}.
\newblock Wiley Series in Probability and Mathematical Statistics. John Wiley
  {\&} Sons, New York, NY, 1977.

\bibitem[JMBO01]{JEON01}
H.~Jeong, S.P. Mason, A.-L. Barab\'asi, and Z.N. Oltvai.
\newblock Lethality and centrality in protein networks.
\newblock {\em Nature}, 411:41--42, 2001.

\bibitem[KRL00]{KRAP00}
P.L. Krapivsky, S.~Redner, and F.~Leyvraz.
\newblock Connectivity of growing random networks.
\newblock {\em Physical Review Letters}, 85:4629--4632, 2000.

\bibitem[KRR{\etalchar{+}}00]{KUMA00b}
R.~Kumar, P.~Raghavan, S.~Rajagopalan, D.~Sivakumar, A.~Tomkins, and E.~Upfal.
\newblock Stochastic models for the web graph.
\newblock In {\em Proceedings of IEEE Symposium on Foundations of Computer
  Science}, pages 57--65, Redondo Beach, Ca., 2000.

\bibitem[LFLW02]{LEVE01c}
M.~Levene, T.I. Fenner, G.~Loizou, and R.~Wheeldon.
\newblock A stochastic model for the evolution of the {W}eb.
\newblock {\em Computer Networks}, 39:277--287, 2002.

\bibitem[MBSA02]{MOSS02}
S.~Mossa, M.~Barth\'{e}l\'{e}my, H.E. Stanley, and L.A.N. Amaral.
\newblock Truncation power law behavior in ``scale-free'' network models due to
  information filtering.
\newblock {\em Physical Review Letters}, 88:138701--1--138701--4, 2002.

\bibitem[New01]{NEWM01}
M.E.J. Newman.
\newblock The structure of scientific collaboration networks.
\newblock {\em Proceedings of the National Academy of Sciences of the United
  States of America}, 98:404--409, 2001.

\bibitem[PFL{\etalchar{+}}02]{PENN02}
D.M. Pennock, G.W. Flake, S.~Lawrence, E.J. Glover, and C.L. Giles.
\newblock Winners don't take all: {C}haracterizing the competition for links on
  the web.
\newblock {\em Proceedings of the National Academy of Sciences of the United
  States of America}, 99:5207--5211, 2002.

\bibitem[Red98]{REDN98}
S.~Redner.
\newblock How popular is your paper? {A}n empirical study of the citation
  distribution.
\newblock {\em European Physical Journal B}, 4:131--134, 1998.

\bibitem[Ros83]{ROSS83}
S.M. Ross.
\newblock {\em Introduction to Stochastic Dynamic Programming}.
\newblock Academic Press, New York, NY, 1983.

\bibitem[Sim55]{SIMO55}
H.A. Simon.
\newblock On a class of skew distribution functions.
\newblock {\em Biometrika}, 42:425--440, 1955.

\bibitem[Sor00]{SORN00}
D.~Sornette.
\newblock {\em Critical Phenomema in the Natural Sciences: Chaos, Fractals,
  Selforganization and Disorder: Concepts and Tools}.
\newblock Springer Series in Synergetics. Springer-Verlag, Berlin, 2000.

\end{thebibliography}
\end{document}